\documentclass[11pt,preprint]{aastex}

\def\ni{\noindent}

\shorttitle{The SMC P-L relations in the {\it IRAC} bands }
\shortauthors{Ngeow \& Kanbur}

\begin{document}

\title{The Mid-Infrared Period-Luminosity Relations for the Small Magellanic Cloud Cepheids Derived from {\it Spitzer} Archival Data}

\author{Chow-Choong Ngeow}
\affil{Graduate Institute of Astronomy, National Central University, Jhongli City, 32001, Taiwan (R.O.C.)}

\and 

\author{Shashi M. Kanbur}
\affil{Department of Physics, State University of New York at Oswego, Oswego, NY 13126}

\begin{abstract}

In this paper we derive the {\it Spitzer IRAC} band period-luminosity (P-L) relations for the Small Magellanic Cloud (SMC) Cepheids, by matching the {\it Spitzer} archival SAGE-SMC data with the OGLE-III SMC Cepheids. We find that the $3.6\mu{\mathrm m}$ and $4.5\mu{\mathrm m}$ band P-L relations can be better described using two P-L relations with a break period at $\log(P)=0.4$: this is consistent with similar results at optical wavelengths for SMC P-L relations. The $5.8\mu{\mathrm m}$ and $8.0\mu{\mathrm m}$ band P-L relations do not extend to sufficiently short periods to enable a similar detection of a slope change at $\log(P)=0.4$. The slopes of the SMC P-L relations, for $\log(P)>0.4$, are consistent with their LMC counterparts that were derived from a similar dataset. They are also in agreement with those obtained from a small sample of Galactic Cepheids with parallax measurements. 

\end{abstract}


\section{Introduction}

A precise and accurate measurement of the Hubble constant at the $\sim2\%$ levels is important for modern precision cosmology \citep[for example, see][]{teg04,hu05,mac06,jac07,oll07,gre09,kom10,lar10,mor10}. This will be feasible after the launch of the {\it James Webb Space Telescope (JWST)}, which will be capable of routinely observing Cepheids in distant galaxies \citep{rie09,fre10} at near- and mid-infrared wavelengths. In preparation for future distance scale work, it is necessary to derive the Cepheid Period-Luminosity relations (hereafter P-L relations, also known as the Leavitt Law) at mid-infrared ($\sim3\mu{\mathrm m}$ to $\sim8\mu{\mathrm m}$) wavelengths. Some advantages and potential problems of the mid-infrared P-L relations were summarized, for example, in \citet[][and reference therein]{nge09a}. For example, the contribution of extinction to the error budget of the Hubble constant becomes negligible in these bands. Motivated by this, the {\it Spitzer IRAC} band (hereafter {\it IRAC} band) P-L relations were derived in \citet{nge08} and \citet{fre08}, by matching the known Cepheids in the Large Magellanic Cloud (LMC) to the archival SAGE dataset \citep[Surveying the Agents of a Galaxy's Evolution,][]{mei06}. The {\it IRAC} band P-L relations were further refined in \citet{nge09} and \citet{mad09}, respectively, by using the SAGE Epoch 1 and 2 data. Besides the LMC Cepheids, the {\it IRAC} band P-L relations were also derived from a sample of Galactic Cepheids that possess independent distance estimates from literature \citep{mar10}.

The main purpose of this Paper is to extend the {\it IRAC} band P-L relations to the Small Magellanic Cloud (SMC), which has a much lower metallicity than both the LMC and our Galaxy. The slopes of the P-L relations in the {\it IRAC} bands are expected to be insensitive to metallicity \citep{fre08}. An empirical determination of the {\it IRAC} band P-L relation derived from SMC Cepheids will provide a critical test of this assumption.

\section{The Data}

Recently, the Optical Gravitational Lensing Experiment has released a catalog of SMC Cepheids from its third phase of operation \citep[OGLE-III,][]{sos10}. This contains $2626$ fundamental mode Cepheids. To derive the SMC {\it IRAC} band P-L relations, we matched the OGLE-III SMC Cepheids to the publicly released Epoch 1 SAGE-SMC \citep[SAGE-SMC: Surveying the Agents of Galaxy Evolution in the Tidally-Disrupted, Low-Metallicity Small Magellanic Cloud,][]{gor10} {\it IRAC} band data (version $S18.0.2$). To be consistent with previous studies \citep{nge08,nge09}, we only use the SAGE-SMC Archive data (hereafter SAGE-SMC data), which contains $\sim1.28$ millions sources. 

We matched the SAGE-SMC sources to the OGLE-III SMC Cepheids by using a search radius of $2''$. The number of matched SAGE-SMC sources and the mean separation ($\Delta$) from the input OGLE-III SMC Cepheids in each {\it IRAC} bands are summarized in Table \ref{separation}. From this table, it can be seen that more than $95\%$ of the matched SAGE-SMC sources are located within $1''$, or within one pixel in the {\it IRAC} band images \citep[the pixel scale for {\it IRAC} instrument is $\sim1.2''/$pixel, see][]{faz04}, from the OGLE-III SMC Cepheids. Extinction is ignored in this Paper, because it is negligible in the {\it IRAC} bands \citep{fre08,nge09}.

\begin{deluxetable}{lcccc}
\tabletypesize{\scriptsize}
\tablecaption{Number and mean separation of the matched SAGE-SMC sources. \label{separation}}
\tablewidth{0pt}
\tablehead{
\colhead{Band} &
\colhead{$N_{\mathrm{match}}$} &
\colhead{$<\Delta>$\tablenotemark{a}} &
\colhead{Std. Dev.\tablenotemark{b}} &
\colhead{Fraction within $1''$\tablenotemark{c}}
}
\startdata
$3.6\mu{\mathrm m}$ & 2567 & 0.263 & 0.214 & 98.75\% \\
$4.5\mu{\mathrm m}$ & 2545 & 0.264 & 0.216 & 98.70\% \\
$5.8\mu{\mathrm m}$ &  699 & 0.265 & 0.271 & 97.00\% \\
$8.0\mu{\mathrm m}$ &  404 & 0.284 & 0.290 & 96.29\%
\enddata
\tablenotetext{a}{$\Delta$ is the separation, in arcsecond, between the matched SAGE-SMC sources and the OGLE-III SMC Cepheids.} 
\tablenotetext{b}{The standard deviation of the mean.}
\tablenotetext{c}{The fraction of matched SAGE-SMC sources within $1''$ radius from the OGLE-III SMC Cepheids.}
%\tablecomments{}
\end{deluxetable}

\section{The SMC P-L Relations in the {\it IRAC} Bands}

\begin{figure*}
\plottwo{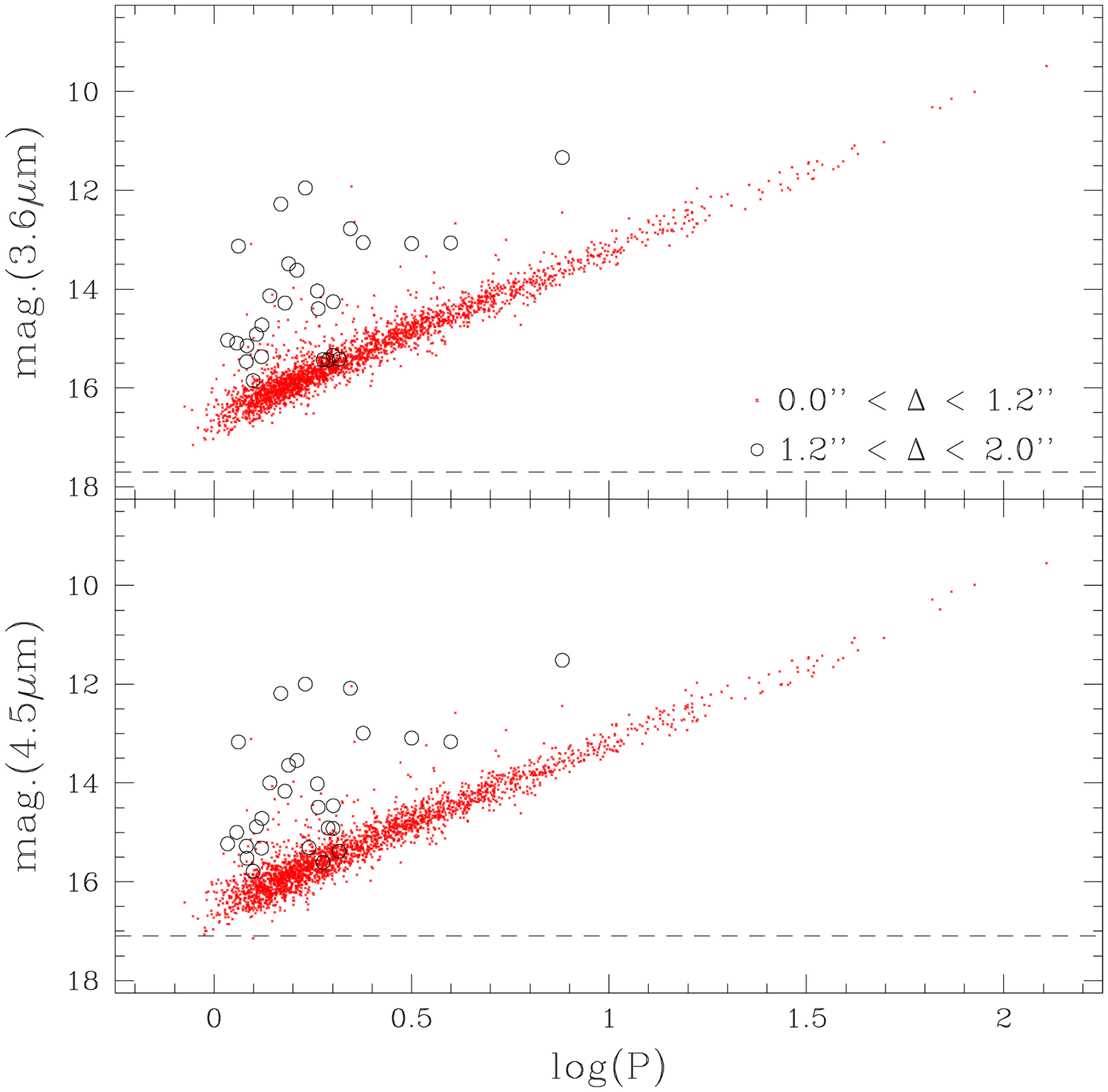}{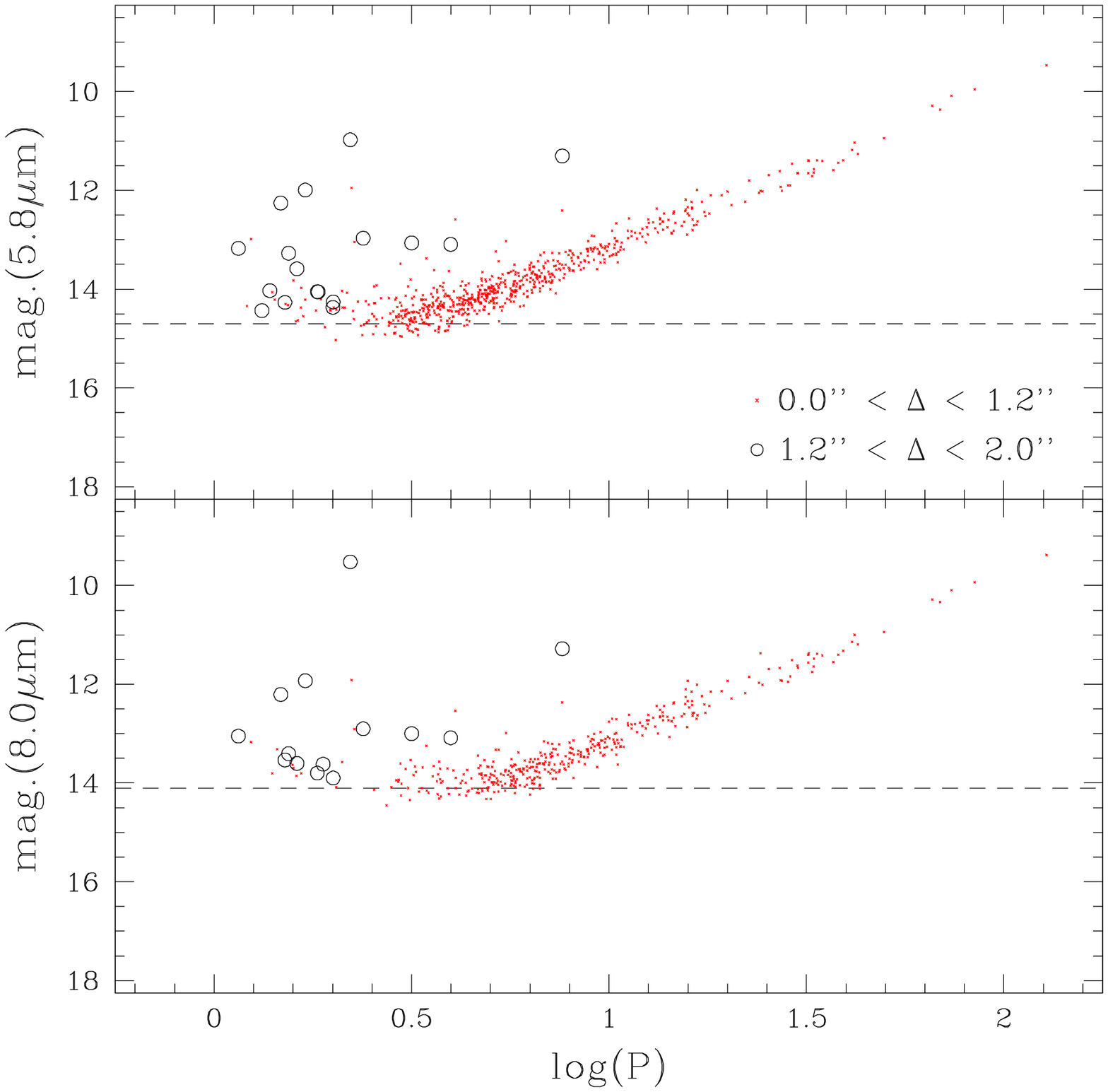}
\caption{The {\it IRAC} band P-L relations for all the matched SAGE-SMC sources to the OGLE-III SMC Cepheid, using a search radius of $2''$. The data were separated in two groups: one group with separations ($\Delta$) between the Cepheid and matched sources that are greater than $1.2''$ (open circles), and another group with $\Delta$ smaller than $1.2''$ (small dots). The dashed horizontal lines are the faint limits in the respected bands, adopted from the SAGE-SMC document (see {\tt http://data.spitzer.caltech.edu/popular/ sage-smc/20091117\_enhanced/documents/sage-smc\_delivery\_nov09.pdf}). $P$ is the pulsating period for Cepheids in days. Error bars are omitted for clarity. \label{rawpl}}
\end{figure*} 

\begin{figure*}
%\plottwo{fig2a.eps}{fig2b.eps}
%\plottwo{fig2c.eps}{fig2d.eps}
$\begin{array}{cccc}
\includegraphics[angle=0,scale=0.4]{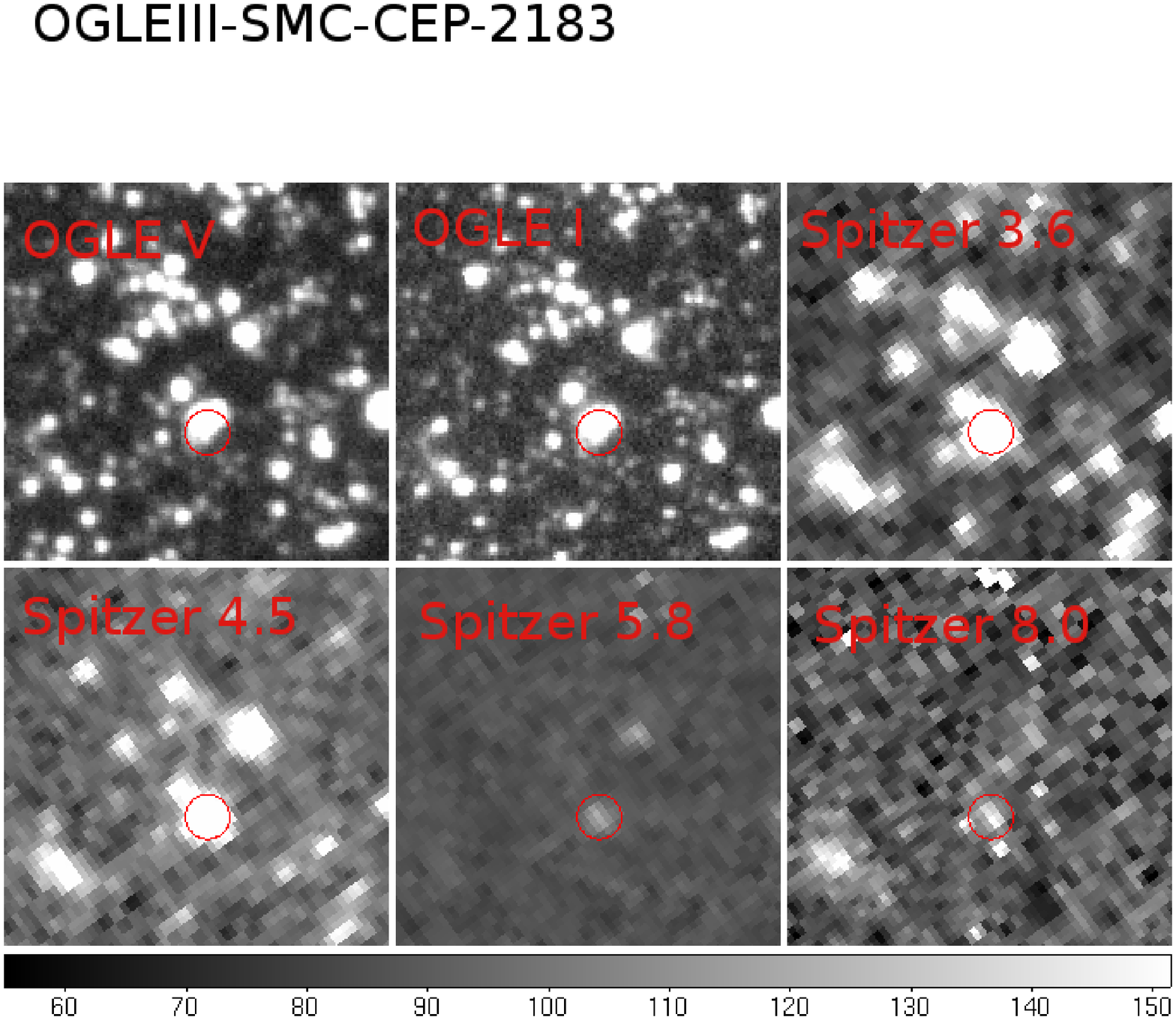} &
\includegraphics[angle=0,scale=0.4]{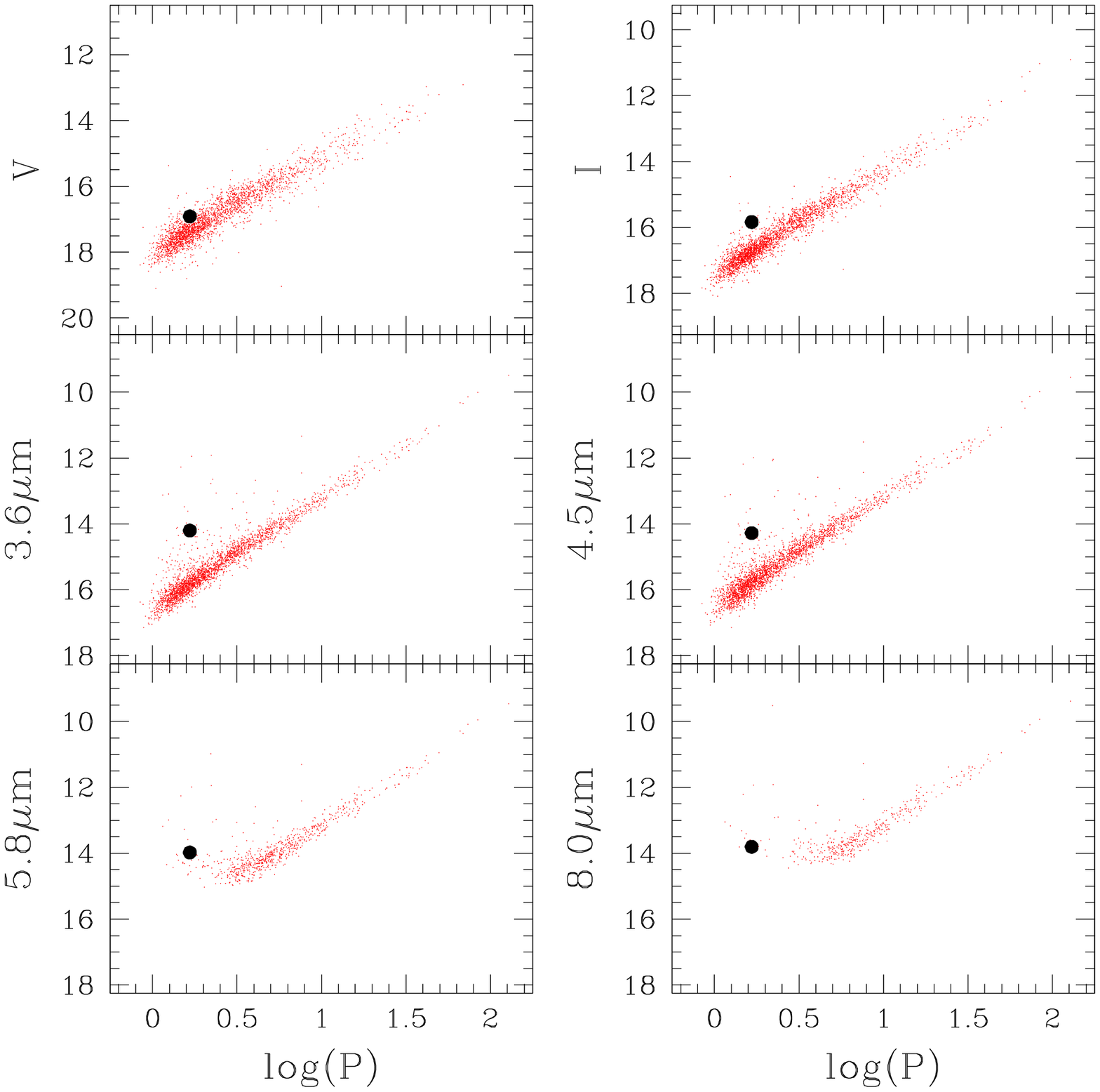} \\
\includegraphics[angle=0,scale=0.4]{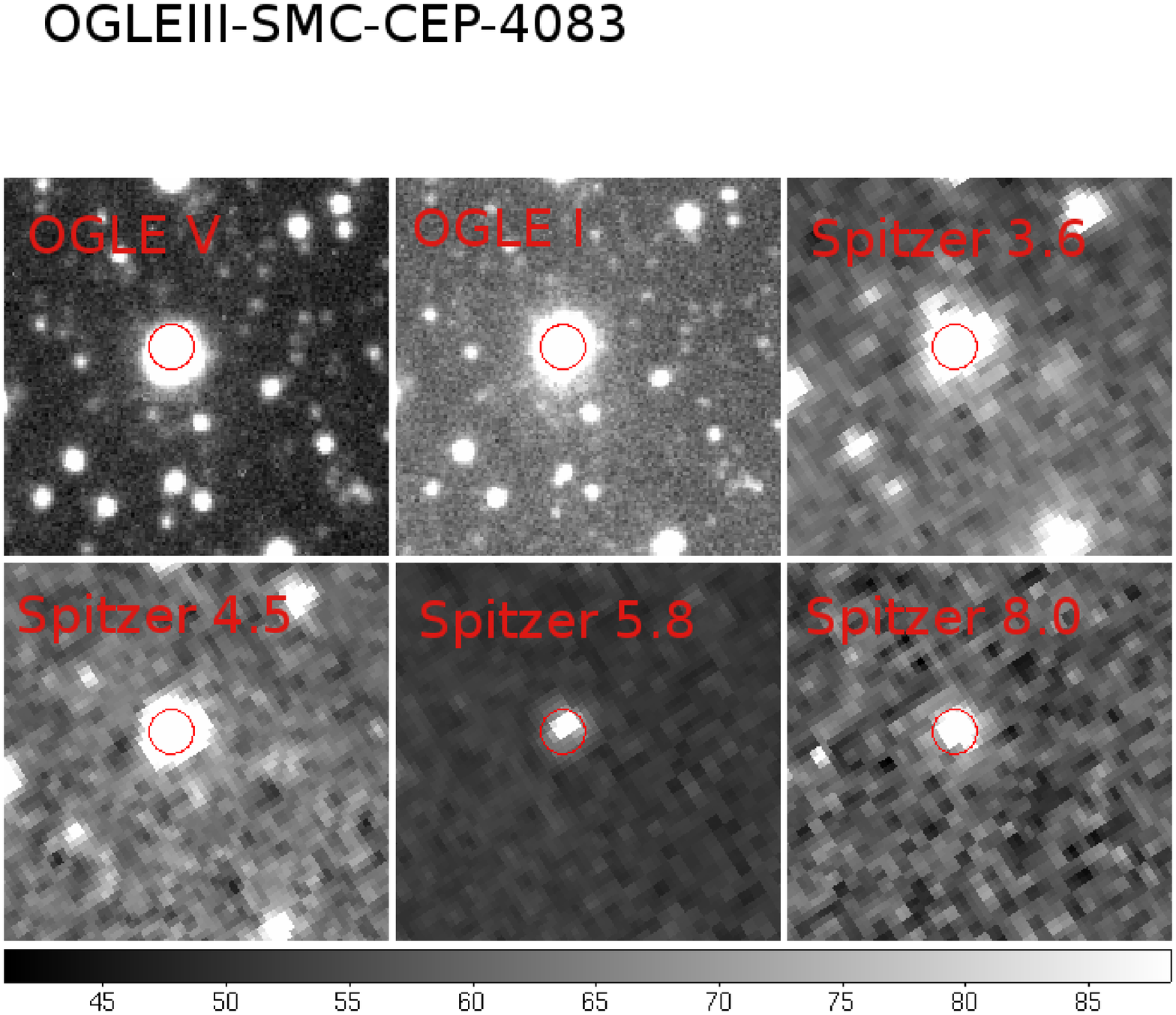} &
\includegraphics[angle=0,scale=0.4]{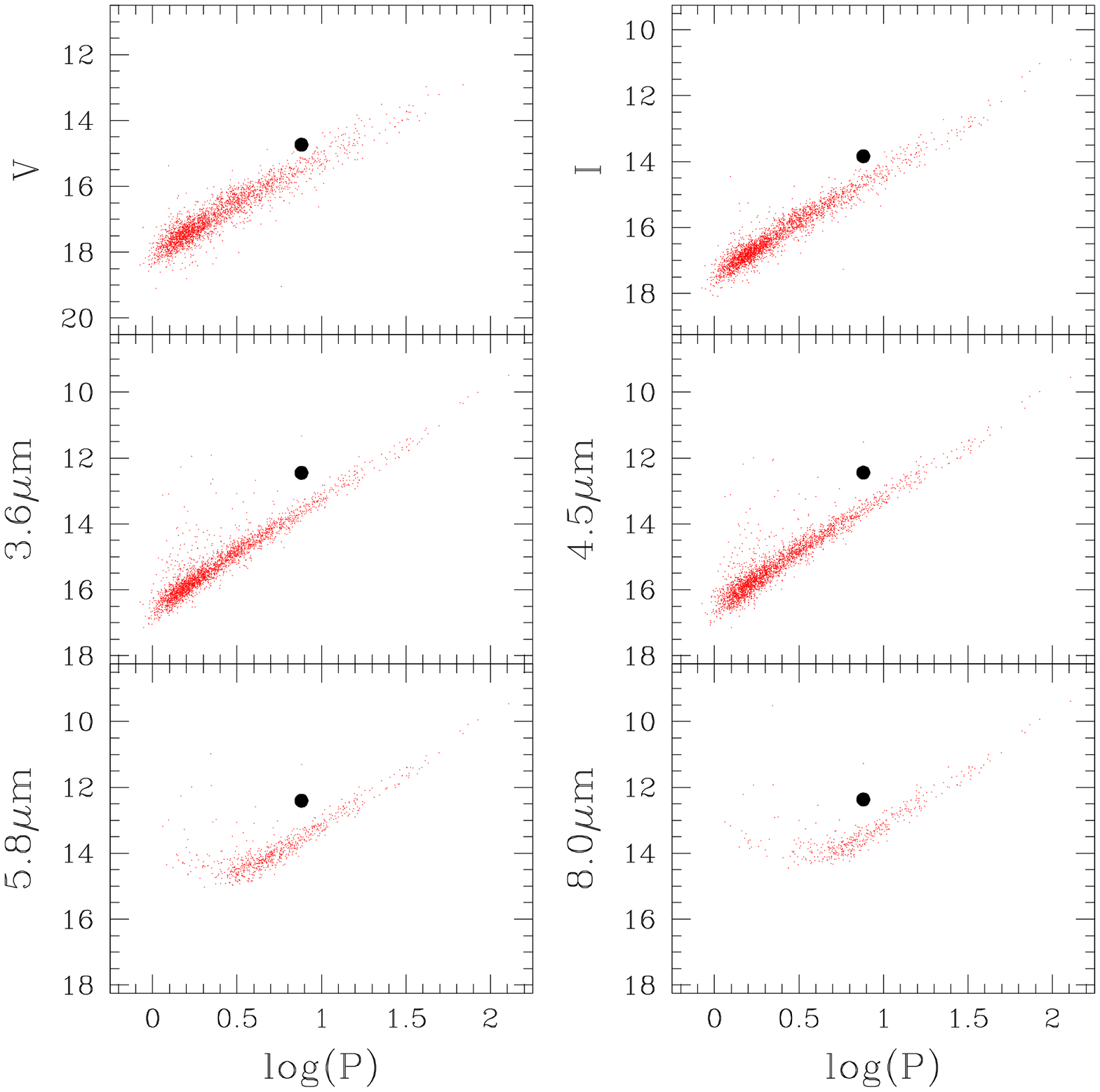} \\
\end{array}$ 
\caption{Two examples of the abnormally bright outliers shown in Figure \ref{rawpl}. The left panels are the postage-stamp images for the two OGLE-III bands and four {\it IRAC} bands. The $VI$ band images were adopted from \citet{uda08}, and the {\it IRAC} images were downloaded from Spitzer Science Center data archive tool, {\tt Leopard}. The circles highlight the location of the Cepheid, which have a radius of $2''$. The large filled circles in right panels show the location of this Cepheid in the respected P-L relations. These two examples do not have any nearby neighboring sources within $1.2''$ from the Cepheid. \label{outlier1}}
\end{figure*} 

\begin{figure*}
%\plottwo{fig3a.eps}{fig3b.eps}
%\plottwo{fig3c.eps}{fig3d.eps}
$\begin{array}{cccc}
\includegraphics[angle=0,scale=0.4]{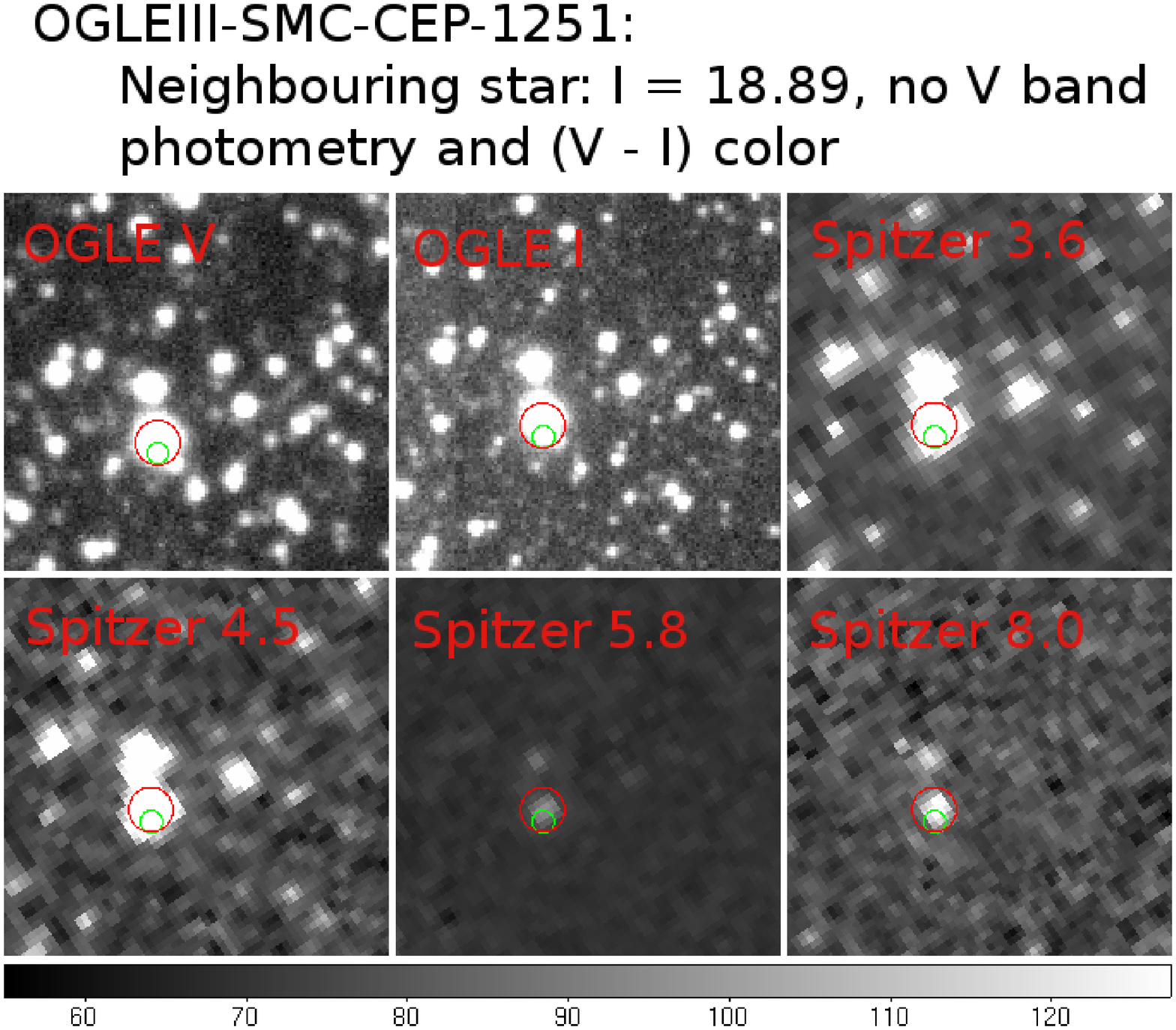} &
\includegraphics[angle=0,scale=0.4]{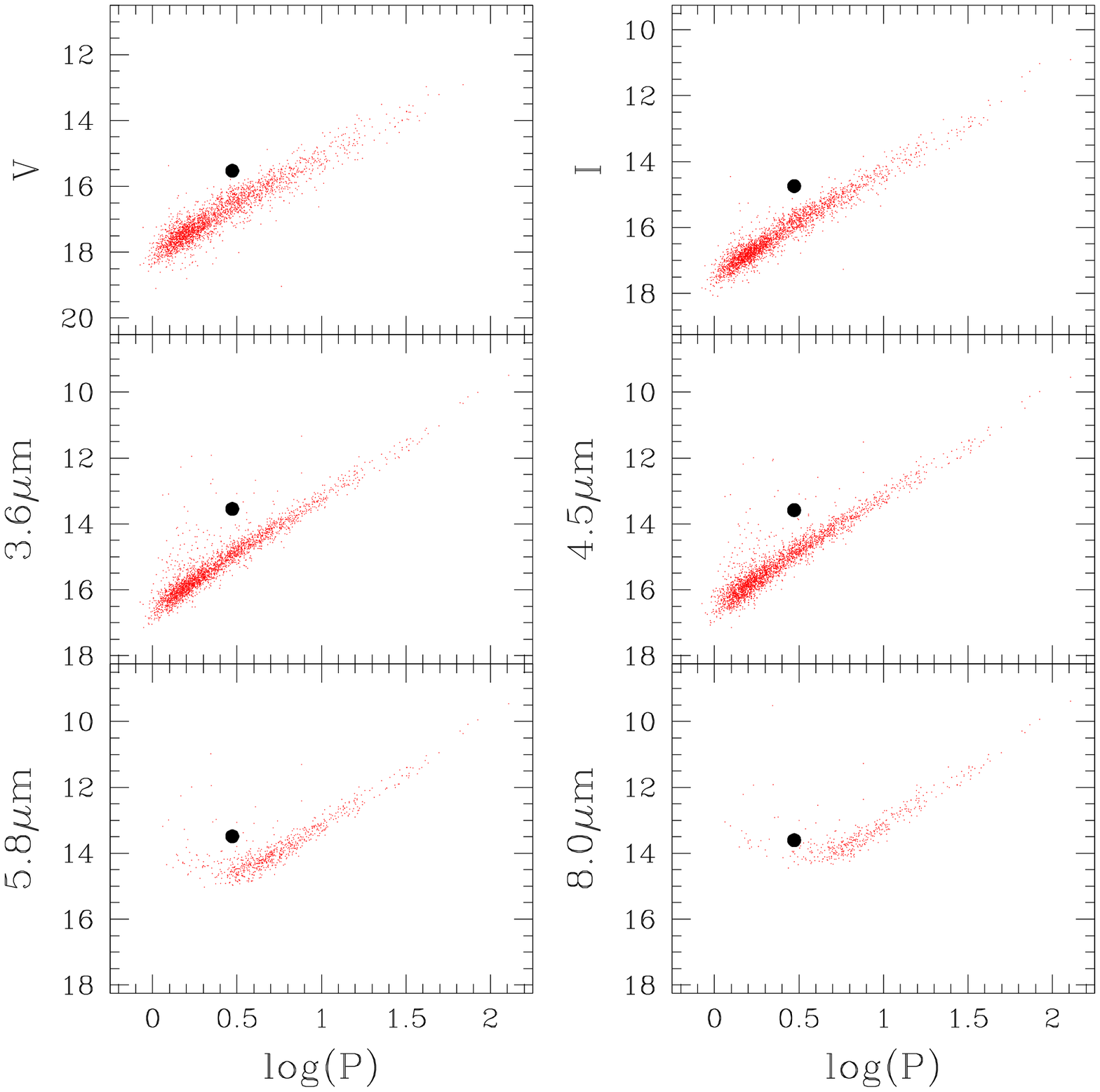} \\
\includegraphics[angle=0,scale=0.4]{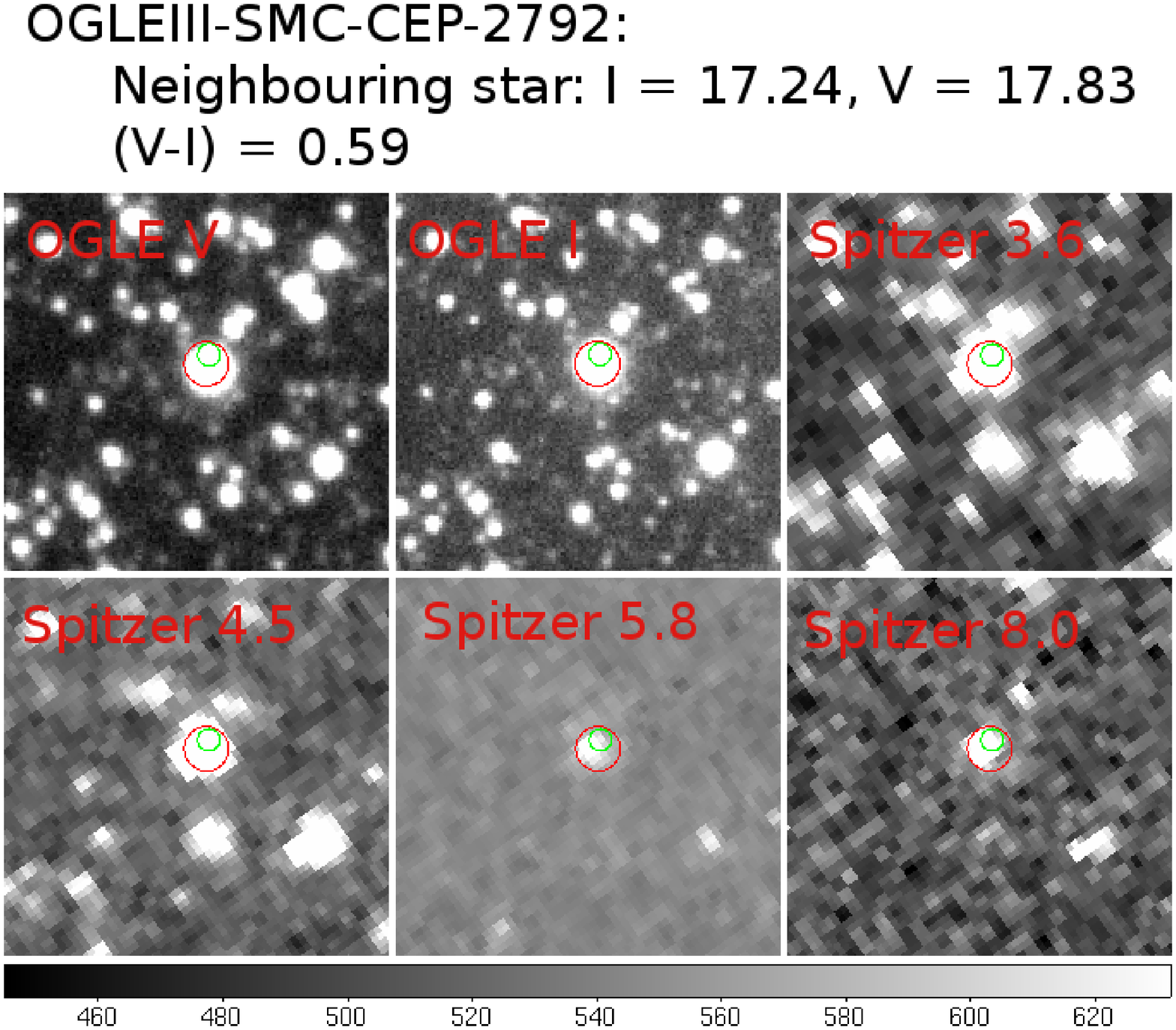} &
\includegraphics[angle=0,scale=0.4]{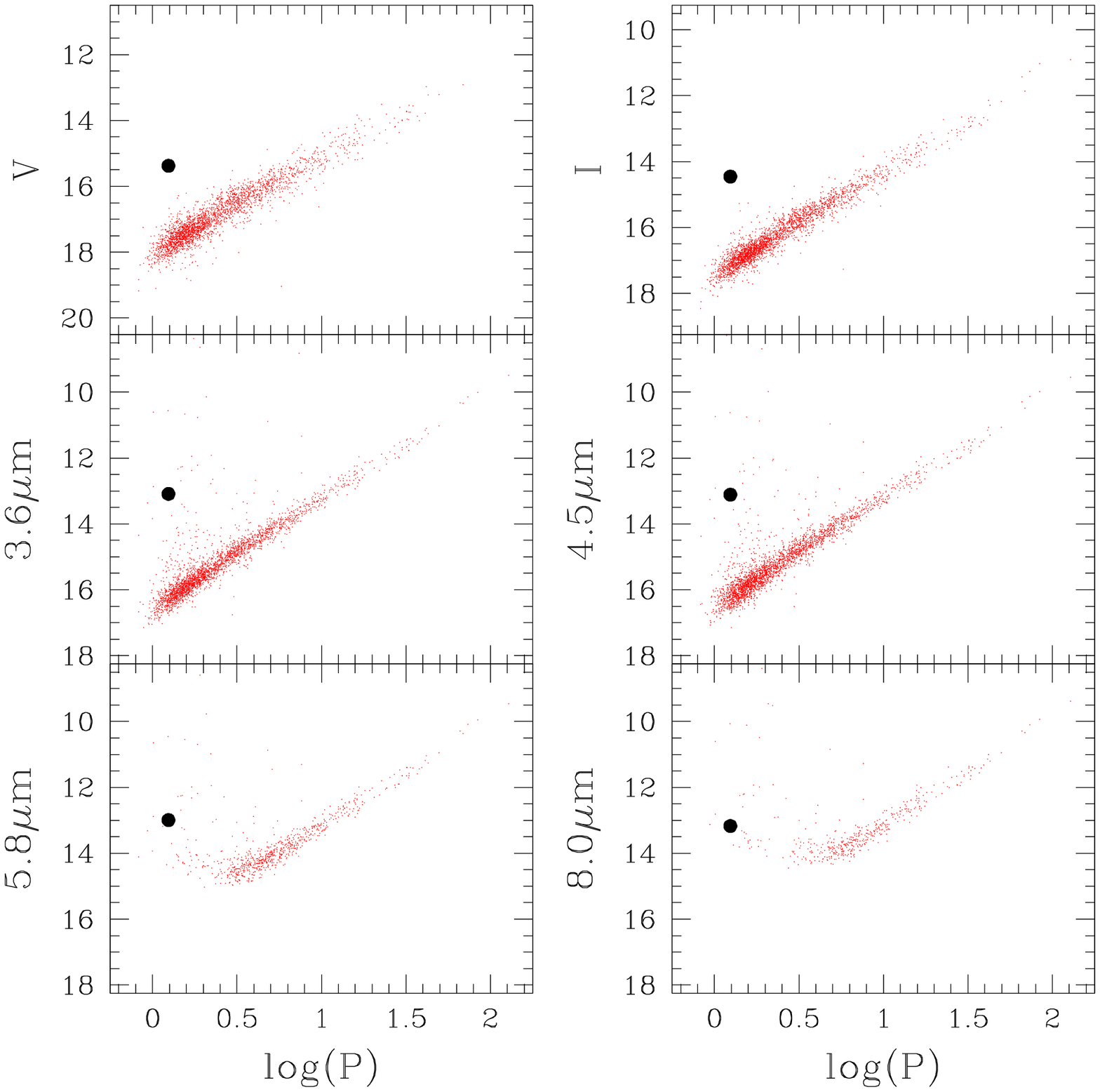} \\
\end{array}$ 
\caption{Same as Figure \ref{outlier1}, but for two Cepheids that have a nearby neighboring source within $1.2''$ from the Cepheid. The location of the neighboring source is indicated with smaller circles (with radius of $1''$) in the left panels. \label{outlier2}}
\end{figure*} 

Figure \ref{rawpl} displays the P-L relation in the four {\it IRAC} bands using all of the matched SAGE-SMC sources. This Figure presents a clear correlation between the {\it IRAC} band magnitudes and the Cepheid pulsation period. To remove the outliers shown in the Figure, we employed an iterative $2.5\sigma$ rejection algorithm when fitting the P-L relations to the data. This is the same procedure used in \citet{nge08} and \citet{nge09}. The abnormally bright outliers shown in Figure \ref{rawpl} could be blended with a nearby red companion. To find out if there are any nearby sources around these outliers, we searched for neighboring sources within $1.2''$ radius from the Cepheids' location, using the OGLE-III SMC photometry map \citep[][which has a pixel scale of $0.26''/$pixel]{uda08}. Only about 20\% of of these outliers were found to have one faint neighboring source around the Cepheids (and few of them have more than one neighboring sources). Figure \ref{outlier1} shows two examples of the outliers that do not have any neighboring source; while Figure \ref{outlier2} presents two examples that display a neighboring source. These abnormally bright outliers could be due to the presence of circumstellar dust envelopes that would cause the observed mid-infrared excess. The circumstellar envelope could be formed from Cepheid mass-loss activity \citep[see, for example,][and reference therein]{nei10}. A detailed investigation of these outliers and their mass-loss activity is beyond the scope of this Paper, but will be addressed in future papers. Nevertheless, it is clear that these outliers should be removed from the sample. 

Figure \ref{rawpl} also suggests that a period cut is needed for the $5.8\mu{\mathrm m}$ and $8.0\mu{\mathrm m}$ band data, as the data approaches the faint magnitude limit for the short period Cepheids. We therefore applied a range of period cuts to the data. Figure \ref{lpcut} shows the slopes of the P-L relations for all four {\it IRAC} bands as a function of period cut. We discuss the implications of the period cuts and the resulting P-L relations in the following sub-sections.

\subsection{The $3.6\mu{\mathrm m}$ \& $4.5\mu{\mathrm m}$ Band P-L Relations}

\begin{figure}
\plotone{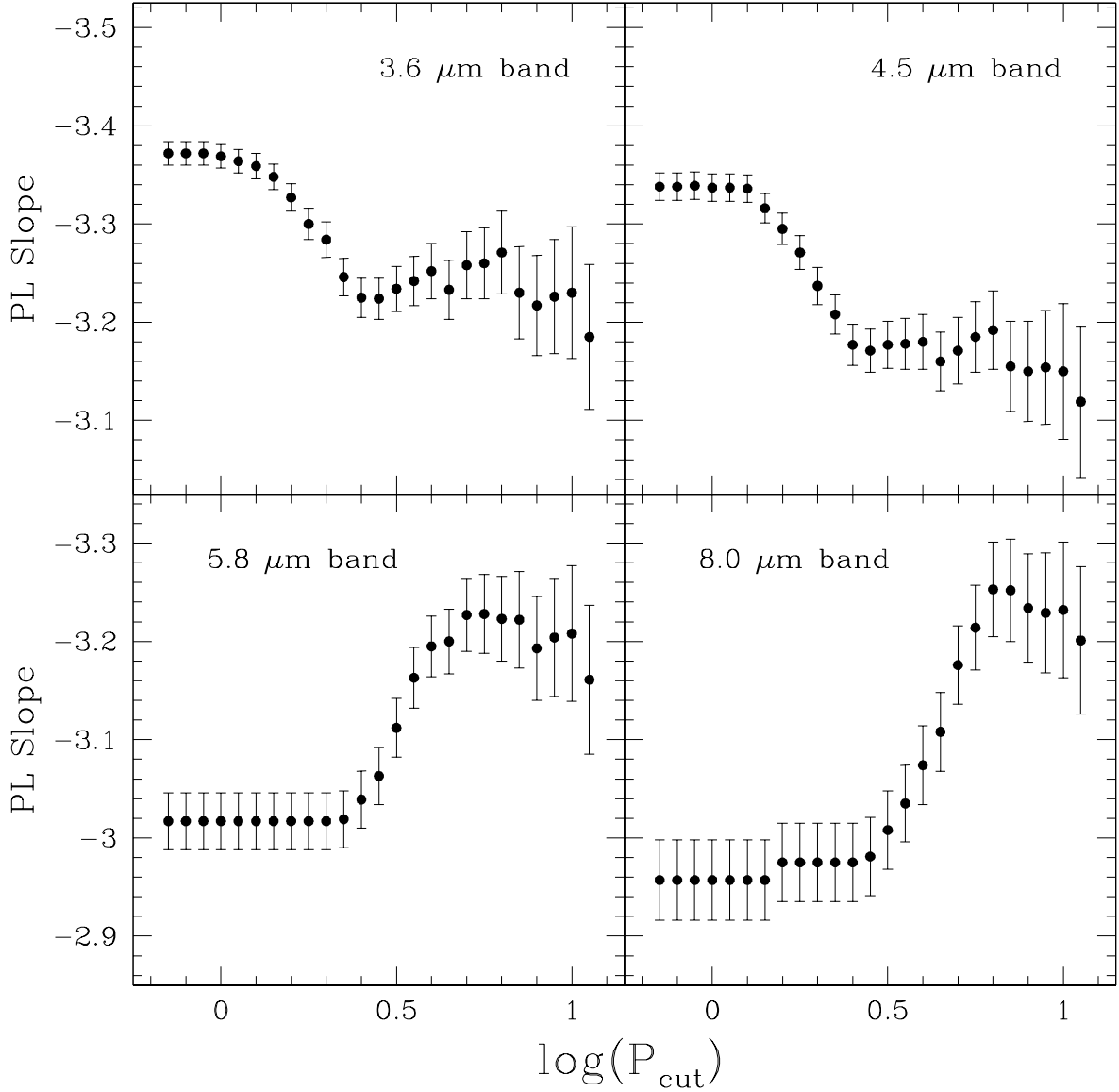}
\caption{The slopes of the P-L relation as a function of adopted period cut ($\log[P_{\mathrm{cut}}]$), where only the data with $\log(P)>\log(P_{\mathrm{cut}})$ were used in deriving the P-L relations. \label{lpcut}}
\end{figure}

Figure \ref{rawpl} implies that period a cut may not be needed for the $3.6\mu{\mathrm m}$ and $4.5\mu{\mathrm m}$ band data. If a period cut is not applied to the data, the resulting P-L relations are: $m_{3.6\mu{\mathrm m}}=-3.372(\pm0.012)\log(P)+16.570(\pm0.006)$ ($N=2378$) and $m_{4.5\mu{\mathrm m}}=-3.338(\pm0.014)\log(P)+16.508(\pm0.007)$ ($N=2389$). However, Figure \ref{lpcut} points to the fact that the slopes of the P-L relations in these two {\it IRAC} bands gradually changes from a steep slope ($\sim -3.35$) to a more shallow slope ($\sim -3.2$) when the the adopted period cut, $\log(P_{\mathrm{cut}})$, is less than $\sim0.4$. In the optical bands, a change of slope for SMC P-L relations at $\log(P)\sim0.4$ has been reported in the literature \citep{bau99,uda99,sha02,san09,sos10}. This slope change is postulated to occur due to evolutionary effects \citep{bec77,bar98}. Therefore, we separated out the $3.6\mu{\mathrm m}$ and $4.5\mu{\mathrm m}$ band data at $\log(P)=0.4$. For Cepheids with $\log(P)<0.4$, the resulting P-L relations are:

\begin{eqnarray}
m_{3.6\mu{\mathrm m}} & = & -3.546(\pm0.049)\log(P)+16.617(\pm0.011), \nonumber \\
 &  & \sigma_{3.6\mu{\mathrm m}}=0.183,\ \ N_{3.6\mu{\mathrm m}} = 1478, \nonumber \\
m_{4.5\mu{\mathrm m}} & = & -3.414(\pm0.058)\log(P)+16.537(\pm0.013), \nonumber \\
 &  & \sigma_{4.5\mu{\mathrm m}}=0.214,\ \ N_{4.5\mu{\mathrm m}} = 1481. \nonumber
\end{eqnarray}

\ni While for Cepheids with $\log(P)>0.4$, the resulting P-L relations are:

\begin{eqnarray}
m_{3.6\mu{\mathrm m}} & = & -3.225(\pm0.020)\log(P)+16.448(\pm0.015), \nonumber \\
 &  & \sigma_{3.6\mu{\mathrm m}}=0.163,\ \ N_{3.6\mu{\mathrm m}} = 906, \nonumber \\
m_{4.5\mu{\mathrm m}} & = & -3.177(\pm0.021)\log(P)+16.371(\pm0.016), \nonumber \\
 &  & \sigma_{4.5\mu{\mathrm m}}=0.172,\ \ N_{4.5\mu{\mathrm m}} = 906. \nonumber
\end{eqnarray}

\begin{figure}
\plotone{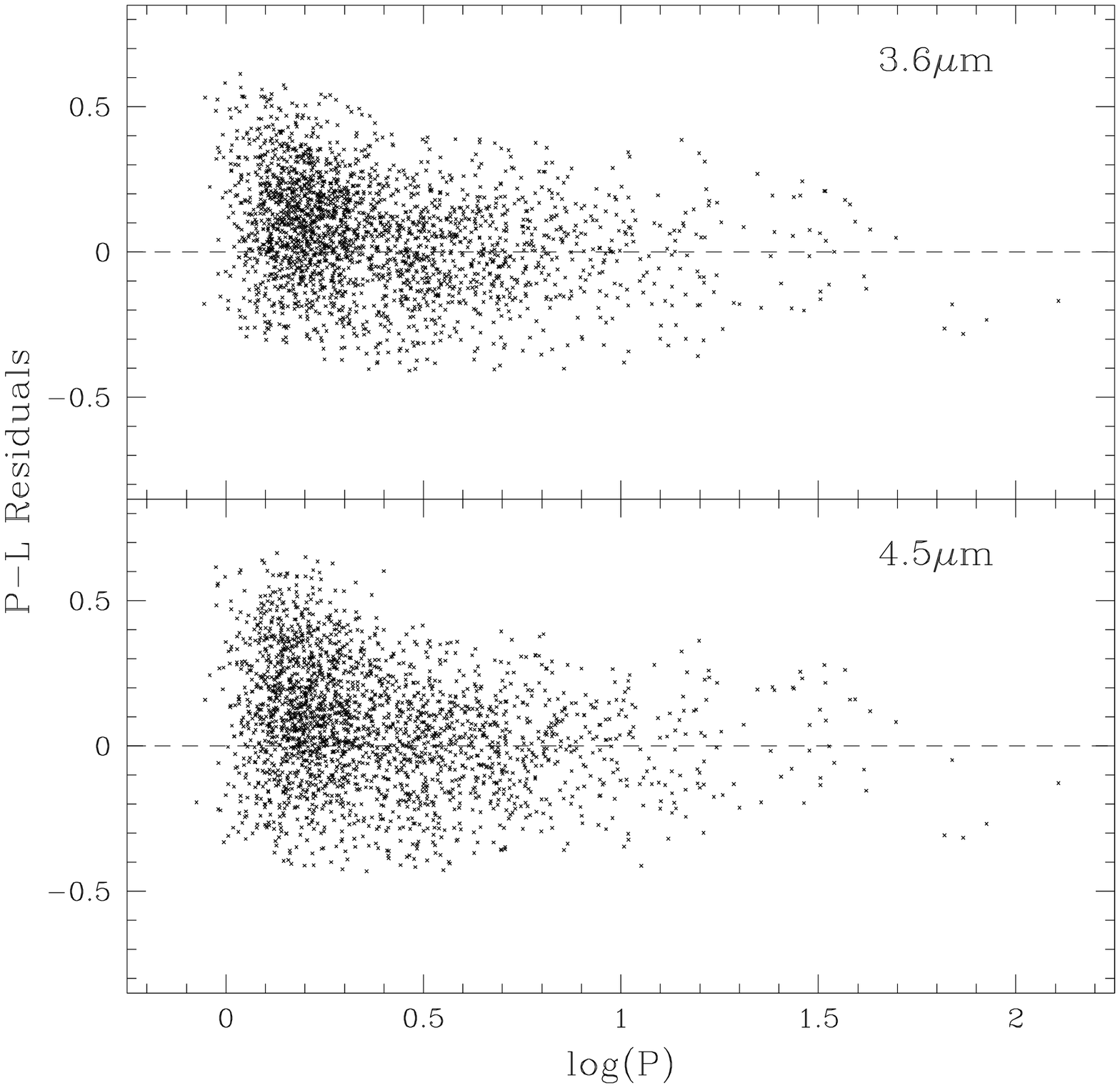}
\caption{Residuals from the fitted P-L relations as a function of period. The P-L relations from the long period ($\log[P]>0.4$) Cepheids were used to fit all data. Upper and lower panels show the $3.6\mu{\mathrm m}$ and $4.5\mu{\mathrm m}$ band residuals, respectively. \label{residual}}
\end{figure}

The difference between the long ($\log[P]>0.4$) and short ($\log[P]<0.4$) period slopes are at the $\sim6\sigma$ and $\sim4\sigma$ level for the $3.6\mu{\mathrm m}$ and $4.5\mu{\mathrm m}$ band P-L relations, respectively. Figure \ref{residual} presents the residuals of the fitted P-L relations as a function of period when the long period P-L relations are used to fit all data. The residuals show a clear trend at the short period end. To further verify that there is a change of slope at $\log(P)=0.4$, we applied the $F$-test \citep[for example, see][]{wei80} to the combined data of the long and short period Cepheids (after removing the outliers). The $F$-test results show that the $3.6\mu{\mathrm m}$ and $4.5\mu{\mathrm m}$ band P-L relations are non-linear at the break period of $\log(P)=0.4$, with $F_{3.6\mu{\mathrm m}}=39.2$ and $F_{4.5\mu{\mathrm m}}=33.1$, respectively\footnote{Recall that for a large number of data points, $F\sim3$ at $95\%$ confident level. Therefore, $F>3$ indicates that the null hypothesis of a single-line regression can be rejected.}. 

For Cepheids with $\log(P)>0.4$, we also tested the non-linear P-L relations at a break period of $\log(P)=1.0$ \citep[which have been found to be non-linear for LMC P-L relations in the optical, see][and reference therein]{kan04,san04,nge05,kan06,nge08b,nge09}. The $F$-test results imply that the P-L relations are linear (with $F_{3.6\mu{\mathrm m}}=0.35$ and $F_{4.5\mu{\mathrm m}}=0.07$). This is consistent  with tests on
the LMC {\it IRAC} band P-L relations \citep{nge08,nge09}.

\subsection{The $5.8\mu{\mathrm m}$ \& $8.0\mu{\mathrm m}$ Band P-L Relations}

It is clear from Figure \ref{rawpl} that the shallow slopes for $5.8\mu{\mathrm m}$ and $8.0\mu{\mathrm m}$ band P-L relations with $\log(P_{\mathrm{cut}})<0.5$, as shown in Figure \ref{lpcut}, are due to the lack of Cepheids at the short period end, where the magnitudes approach the faint limit in these two {\it IRAC} bands. A period cut of $\log(P_{\mathrm{cut}})>0.5$ is necessary to remove the bias due to the incomplete data at the faint magnitude end. Inspecting Figure \ref{lpcut} suggests that the slope of the $5.8\mu{\mathrm m}$ and $8.0\mu{\mathrm m}$ band P-L relations begin to stabilize at $\log(P_{\mathrm{cut}})\sim0.7$ and $\log(P_{\mathrm{cut}})\sim0.8$, respectively. Hence, we have adopted a period at $\log(P_{\mathrm{cut}})=0.7$ and $\log(P_{\mathrm{cut}})=0.8$ when deriving the P-L relations in these two bands. The resulting P-L relations are:

\begin{eqnarray}
m_{5.8\mu{\mathrm m}} & = & -3.227(\pm0.037)\log(P)+16.381(\pm0.038), \nonumber \\
 &  & \sigma_{5.8\mu{\mathrm m}}=0.180,\ \ N_{5.8\mu{\mathrm m}} = 334, \nonumber \\
m_{8.0\mu{\mathrm m}} & = & -3.253(\pm0.048)\log(P)+16.400(\pm0.053), \nonumber \\
 &  & \sigma_{8.0\mu{\mathrm m}}=0.189,\ \ N_{8.0\mu{\mathrm m}} = 227. \nonumber
\end{eqnarray}

The ratios of the number of Cepheids with $\log(P)>1.0$ and $\log(P)<1.0$ are $1:2$ and $1:1$ in the $5.8\mu{\mathrm m}$ and $8.0\mu{\mathrm m}$ band, respectively, after the period cut. Therefore the $F$-test can be applied to the data in order to test for non-linearity at $\log(P)=1.0$. The $F$-test results show that both of the $5.8\mu{\mathrm m}$ and $8.0\mu{\mathrm m}$ band P-L relations are linear, with $F_{5.8\mu{\mathrm m}}=0.06$ and $F_{8.0\mu{\mathrm m}}=0.08$ respectively. This result is also consistent with the LMC results \citep{nge08,nge09}.

\section{Comparison of the P-L Slopes}

In Table \ref{slopecompare}, we compare the {\it IRAC} band P-L slopes currently available for our Galaxy, LMC and SMC. We aggregate the slopes into two groups: one with shallow slopes and another with steep slopes. In general, slopes within a group are consistent with each other. In contrast, the LMC slopes derived in \citet[][the LMC2]{nge09} and the SMC slopes disagree with those in the ``steep slope'' group at the $3\sigma$ level or more. If the {\it IRAC} band P-L slopes were indeed shallow, then the slopes from all three galaxies would be consistent (despite the large uncertainties for GAL3 slopes shown in Table \ref{slopecompare}). This suggests that the metallicity will not strongly affect the {\it IRAC} band P-L slopes. On the other hand, if the Galactic and LMC P-L slopes were steeper (for those listed as GAL1/2 and LMC1 in Table \ref{slopecompare}), our SMC slopes will challenge the assumption that these slopes should be insensitive to metallicity. 

\begin{deluxetable}{lcccccl}
\tabletypesize{\scriptsize}
\tablecaption{Comparison of the P-L slopes in the {\it IRAC} bands. \label{slopecompare}}
\tablewidth{0pt}
\tablehead{
\colhead{Galaxy} &
\colhead{$3.6\mu{\mathrm m}$} & 
\colhead{$4.5\mu{\mathrm m}$} &
\colhead{$5.8\mu{\mathrm m}$} &
\colhead{$8.0\mu{\mathrm m}$} &
\colhead{Reference\tablenotemark{a}} &
\colhead{Note} 
}
\startdata
\cutinhead{Steep Slope} \\
GAL1  & $-3.54\pm0.04$ & $-3.45\pm0.04$ & $-3.51\pm0.03$ & $-3.57\pm0.03$ & (1) & $29$ GAL Cepheids with ``old'' IRSB distances\tablenotemark{b} \\
GAL2  & $-3.47\pm0.06$ & $-3.38\pm0.06$ & $-3.44\pm0.06$ & $-3.46\pm0.06$ & (1) & $28$ GAL Cepheids with ``new'' IRSB distances\tablenotemark{b} \\
LMC1  & $-3.40\pm0.02$ & $-3.35\pm0.02$ & $-3.44\pm0.03$ & $-3.49\pm0.03$ & (2) & $70$ LMC Cepheids \\
\cutinhead{Shallow Slope} \\
GAL3  & $-3.16\pm0.22$ & $-3.06\pm0.23$ & $-3.10\pm0.23$ & $-3.16\pm0.18$ & (1) & $8$ GAL Cepheids with parallax distances\tablenotemark{b} \\
LMC2  & $-3.25\pm0.01$ & $-3.21\pm0.01$ & $-3.18\pm0.02$ & $-3.20\pm0.04$ & (3) & $\sim400$ to $\sim1630$ LMC Cepheids \\
SMC   & $-3.23\pm0.02$ & $-3.18\pm0.02$ & $-3.23\pm0.04$ & $-3.25\pm0.05$ & (4) & $\sim220$ to $\sim900$ SMC Cepheids \\
\enddata
\tablenotetext{a}{Reference: (1) \citet{mar10}; (2) \citet{mad09}; (3) \citet{nge09}; (4) This Paper for Cepheids with $\log(P)>0.4$.}
\tablenotetext{b}{``Old'' and ``new'' distances referred to the old and new projection factor-period relations adopted in IRSB technique, respectively. See \citet{mar10} for more details.}
\end{deluxetable}

Does crowding affect the derivation of the SMC P-L slopes presented in this Paper? We tested this by dividing our SMC Cepheids into two groups: one group for Cepheids located in the central ``bar'' region (with higher stellar density), and another group for Cepheids located on the outskirts of the ``bar'' region. The derived P-L relations from these two groups are consistent with each other. We performed another test by searching the nearby companion sources within $1.2''$ of the respected Cepheids, using the OGLE-III SMC photometry map mentioned previously. The Cepheid samples were grouped into two groups that either did or did not contain the nearby sources. Again, the slopes derived from these two groups are in good agreement. We therefore believe crowding does not seriously affect our results. 

The discrepancy between the two LMC slopes given in Table \ref{slopecompare} may be due, in part, to the different samples used in both studies. \citet{mad09} derived the P-L relations using 70 LMC Cepheids drawn from \citet{per04}. Most of them had periods longer than $10$ days. In contrast, the LMC P-L relations presented in \citet{nge09} were derived based on the $\sim1800$ OGLE-III LMC Cepheids \citep{sos08}, which have a distribution peaking at $\log(P)\sim0.5$ \citep{sos10}. \citet{nei10} showed that the LMC {\it IRAC} band P-L slopes, based on the OGLE-III Cepheids, became comparable to the slopes given in \citet{mad09} when only the long period ($\log[P]>1$) Cepheids were used. 

For the Galactic {\it IRAC} band P-L relations, the P-L slopes derived from the infrared surface brightness (IRSB) technique (GAL1 and GAL2 in Table \ref{slopecompare}) were steeper than the slopes derived from Cepheids that have parallax measurements (GAL3 in Table \ref{slopecompare}). The large uncertainties of the GAL3 P-L slopes \citep[due to the combination of a small number of data points in the sample and less precise photometry, see][]{mar10} cause all three sets of Galactic P-L slopes to be consistent with each other. It is worthwhile to note that the theoretical P-L slopes derived in \citet{mar10} agree well with the GAL3 P-L slopes. If the uncertainties of the P-L slopes derived from the parallax distances are the same as those from the IRSB techniques, then the slopes between them will disagree at the $\sim 3.5\sigma$ level. \citet{mar10} suggested that the projection factor-period relation used in IRSB technique may need further refinement.

Finally, we have derived the {\it AKARI N3} band (at $\sim3\mu{\mathrm m}$) P-L relation for the LMC Cepheids, which is independent of the SAGE data, in \citet{nge10}. The slope of the $N3$ band P-L relation is $-3.25\pm0.05$. This agrees well with the $3.6\mu{\mathrm m}$ band P-L slopes in Table \ref{slopecompare} for the slopes in the ``Shallow Slope'' group.

\section{Conclusion}

In this Paper we have derived SMC P-L relations in the {\it IRAC} bands, by matching the archival SAGE-SMC data to the latest SMC Cepheid catalog from OGLE-III. We have found, for the first time, that the SMC P-L relations show a change of slope at $\log(P)=0.4$ in the $3.6\mu{\mathrm m}$ and $4.5\mu{\mathrm m}$ band, similar to their optical counterparts. Due to the incompleteness at the short period end, such a change of slope cannot be confirmed to exist in the $5.8\mu{\mathrm m}$ and $8.0\mu{\mathrm m}$ bands. Future observations with {\it JWST} may help to determine the change of slope in these two bands. 

The slopes of the SMC {\it IRAC} band P-L relations were found to be around $-3.2$, which is consistent with the LMC slopes found in \citet{nge09} and the Galactic P-L relations derived from Cepheids with parallax measurements \citep{mar10}. This indicates that the slopes of the P-L relations in the {\it IRAC} bands are insensitive to metallicity. However, the SMC slopes disagree with the steeper slopes for the LMC and Galactic counterparts from \citet{mad09} and those from the IRSB technique, respectively. Future observations such as the parallax measurements from {\it Gaia} for Galactic Cepheids and/or {\it JWST} observations of Magellanic Cloud Cepheids may help to resolve this discrepancy. 

\acknowledgments

CCN thanks the funding from National Science Council (of Taiwan) under the contract NSC 98-2112-M-008-013-MY3. We would like to thank the referee for helpful comments to improve the manuscript. We would also like to thank Hilding Neilson and Nancy Evans for useful discussions. This work is based [in part] on observations made with the Spitzer Space Telescope, which is operated by the Jet Propulsion Laboratory, California Institute of Technology under a contract with NASA.

\end{document}